\documentclass[11pt,english,floatfix,superscriptaddress,aps,preprint,showpacs]{revtex4}

\usepackage{color}

\usepackage[english]{babel}

\usepackage{amsmath,amssymb,amsfonts}

\usepackage{graphicx}

\begin{document}

\title{Three-dimensional Dirac oscillator in a thermal bath}

\author{M. H. Pacheco}
\email{pacheco@fisica.ufc.br}
\affiliation{Universidade Federal do Cariri (UFCA), Cidade Universit\'{a}ria , Campus
do Cariri, 63000-000, Juazeiro do Norte, Cear\'{a}, Brazil}

\author{R. V. Maluf}
\email{r.v.maluf@fisica.ufc.br}
\affiliation{Universidade Federal do Cear\'{a} (UFC), Departamento de F\'{i}sica, Campus
do Pici, C.P. 6030, 60455-760, Fortaleza , Cear\'{a}, Brazil}

\author{C. A. S. Almeida}
\email{carlos@fisica.ufc.br}
\affiliation{Universidade Federal do Cear\'{a} (UFC), Departamento de F\'{i}sica, Campus
do Pici, C.P. 6030, 60455-760, Fortaleza , Cear\'{a}, Brazil}

\author{R. R. Landim}
\email{renan@fisica.ufc.br}
\affiliation{Universidade Federal do Cear\'{a} (UFC), Departamento de F\'{i}sica, Campus
do Pici, C.P. 6030, 60455-760, Fortaleza , Cear\'{a}, Brazil}




\begin{abstract}The thermal properties of the three-dimensional Dirac oscillator are considered.
The canonical partition function is determined, and the high-temperature limit is assessed. The degeneracy of energy levels and their physical implications on the main thermodynamic functions are analyzed, revealing that these functions assume values greater than the one-dimensional case. So that at high temperatures, the limit value of the specific heat is three times bigger.
\end{abstract}


\maketitle

\section{Introduction}

The harmonic oscillator potential for relativistic spin 1/2 was first
introduced by Ito et al. \cite{Ito1967} and has attracted considerable
interest in the last years. Moshinsky and Szczepaniak named it Dirac
oscillator and renewed to a great extent the interest in the topic
\cite{Moshinsky1989}. Besides intrinsic mathematical interest, the
study of the Dirac oscillator has drawn much attention because of its
various physical applications. For instance, Moshinsky and Szczepaniak
\cite{Moshinsky1989} showed that, in the non-relativistic limit,
its becomes a harmonic oscillator with a strong spin-orbit coupling
term plus a constant term. Moreno and Zentella \cite{Moreno1989}
showed that an exact Foldy-Wouthuysen transformation could be performed
in the Dirac oscillator. Also, Quesne and Moshinsky \cite{Quesne1990}
studied its symmetry Lie algebra explicitly. After, Benitez et al.
\cite{Bentez1990} found the complete energy spectrum and the corresponding
eigenfunction of the Dirac oscillator. They found the electromagnetic
potential associated with the Dirac oscillator, and showed that this
exactly soluble problem has a hidden supersymmetry, which is responsible
for the special properties of its energy spectrum. They also calculated
the related superpotential and discussed the implications of this
supersymmetry on the stability of the Dirac sea.

Some time later, Nogami and Toyama \cite{Nogami1996} and Toyama et
al. \cite{Toyama1997} have studied the behavior of wave packets of
the Dirac oscillator in the $(1+1)$ dimensional Dirac representation.
The aim of these authors was to study wave packets that represent
coherent states. This reduction of the dimension was brought as an
attempt to get rid of spin effects and to concentrate on the relativistic
effects. Also, Rozmej and Arvieu \cite{Rozmej1999} have shown a very
interesting analogy between the relativistic Dirac oscillator and
the Jaynes-Cummings model. They showed that the strong spin-orbit
coupling of the Dirac oscillator produces an entanglement of the spin
with the orbital motion similar to that is observed in quantum optic
models. More recently, some attention has been given to the models
of noncommutative quantum mechanics and the space noncommutativity
version of the Dirac oscillator was studied in Refs. \cite{Mirza2004,Maluf2011}.  Also, regarding the Aharonov-Bohm effect, the Dirac oscillator in noncommutative spaces was considered by Hassanabadi et al. \cite{Hassanababi}. The authors then calculated the exact energy levels and the corresponding eigenfunctions by means of the Nikiforov-Uvarov method. Besides, the Dirac oscillator has also been discussed in connection with the $\kappa$-deformed Poincar\'{e}-Hopf algebra by Andrade et al. \cite{Edilberto}.

The study of the thermal properties of the one-dimensional Dirac oscillator was carried out by some of us, some years ago \cite{Pacheco2003}. We applied a numerical method based on the Euler-MacLaurin formula to calculate the associated partition function and hence obtain the thermodynamic functions, such as the free energy, the mean energy and the heat capacity. Since then, several studies have been made in connection with our initial results \cite{Boumali2007,Boumali2013,Hassanabadi2014,Boumali2014}. On the other hand, the method was initially used for the study of one-dimensional systems  which have non-degenerate energy spectrum. To the best of our knowledge, this approach has not been extended to three-dimensional systems with degeneracies of the eigenvalues. The present paper has as its main goal to extend the work of \cite{Pacheco2003} and to consider the full $3D$ Dirac oscillator. Contrary to one-dimensional case we expect to observe new physical effects due to the presence of a large spin-orbit coupling  and the existence of degeneracy in the energy spectrum.

This work is organized as follows. In Sec. \ref{sec:thermalProperties},
we make a brief review of the Dirac oscillator in a thermal bath, discussing the degeneracy of the stationary states and present our main results on the thermodynamic properties of the model. Furthermore, we comment on the connection of our results with others in the literature. Finally, we present our conclusions and final remarks in Sec. \ref{sec:Conclusions}.

\section{Thermal properties of the 3D Dirac oscillator\label{sec:thermalProperties}}

Before discussing the thermal properties of the Dirac oscillator,
let us first briefly recall the main results on this system. The three-dimensional
Dirac equation for a free spin $1/2$ particle is
\begin{equation}
\left(c\boldsymbol{\alpha}\cdot\boldsymbol{p}+\beta mc^{2}\right)\psi=E\psi,
\end{equation}
where $\boldsymbol{p}=-i\hbar\nabla$, $\boldsymbol{\alpha}$, $\beta$
are usual Dirac matrices, $m$ is the rest mass of the particle and
$c$ is the speed of light. The Dirac oscillator can be obtained through
the non-minimal substitution $\boldsymbol{p}\rightarrow\boldsymbol{p}-im\omega\beta\boldsymbol{r}$,
which is linear in both the coordinate and the momentum. This leads
to the Hamiltonian
\begin{equation}
H=c\boldsymbol{\alpha}\cdot\left(\boldsymbol{p}-im\omega\beta\boldsymbol{r}\right)+\beta mc^{2},
\end{equation}
with $\omega>0$ being the constant oscillator frequency. The non-relativistic
limit of the Dirac oscillator reproduces the usual harmonic oscillator
added by a very large spin-orbit coupling plus a constant term. 

The complete solution for the Dirac oscillator was first obtained
by Moshinsky and Szczepaniak \cite{Moshinsky1989} and the corresponding
energy levels can be expressed as
\begin{equation}
E_{Nlj}^{2}=\left\{ \begin{array}{lll}
m^{2}c^{4}+(2N-2j+1)\hbar\omega mc^{2} & \mbox{if} & j=l+\frac{1}{2}\\
m^{2}c^{4}+(2N+2j+3)\hbar\omega mc^{2} & \mbox{if} & j=l-\frac{1}{2}
\end{array}\right.,\label{energy}
\end{equation}
where $j$, $l$ and $N$ are the total angular momentum, the orbital
angular momentum and the principal quantum numbers, respectively.
These expressions can be rewritten in another manner as
\begin{equation}
E_{Nlj}^{2}=\left\{ \begin{array}{lll}
m^{2}c^{4}+4n\hbar\omega mc^{2} & \mbox{if} & j=l+\frac{1}{2}\\
m^{2}c^{4}+(4n+4l+2)\hbar\omega mc^{2} & \mbox{if} & j=l-\frac{1}{2}
\end{array}\right.,\label{energy-1}
\end{equation}
such that $N=2n+l$ while $n=0,1,2,\ldots$ is the radial quantum
number \cite{Davydov}. We see from Eqs. \eqref{energy}-\eqref{energy-1}
that the stationary states exhibit a great deal of degeneracy. For
instance, the energy levels with $j=l+\frac{1}{2}$ depend only on
the values of $n$. Since $l$ can take any integer value, the states
are infinitely degenerate. In particular, for $j=\frac{1}{2}$ $(l=0)$
all states are two-fold degenerate (with fixed $n$). Further, when
$j=l-\frac{1}{2}$ one has that $l\geqslant1$ and the energy depends
on the combination of quantum numbers $n+l=k$, but now the degeneracy
remains finite, increasing with the $k$ value.

To carry out our analysis on the thermodynamics of the Dirac oscillator,
we will restrict ourselves to stationary states of positive energy
and whose degeneracy remains finite. The reason for this is twofold.
First, the Hamiltonian for the Dirac oscillator admits an exact Foldy-Wouthuysen
transformation (FWT), such that the positive- and negative-energy
solutions never mix, leaving the Dirac sea stable \cite{Martinez1991,Martinez1995}.
Consequently, we can assume that only particles with positive energy
are available in order to set up a thermodynamic ensemble. Second,
the solutions with infinite degeneracy do not correspond to physical
states since there is not Lorentz finite representation for them.

Now let us write the main object of our interest, the canonical partition
function $Z$ at finite temperature $T$ . For the positive energy
levels with $j=l-\frac{1}{2}$, the following equation for $Z$ holds:
\begin{equation}
Z=\sum_{k}\Omega\left(E_{k}\right)\exp\left(-\beta E_{k}\right),
\end{equation}
where $\beta=1/\kappa_{B}T$ , $\kappa_{B}$ is the Boltzmann constant, $\Omega\left(E_{k}\right)$
is the degree of degeneracy for the energy level $E_{k}$ defined
as
\begin{equation}
E_{k}=mc^{2}\sqrt{1+\left(4k+2\right)\xi},
\end{equation}
with $k=n+l$, $k\geqslant1$ and $\xi=\hslash\omega/mc^{2}$ being
a dimensionless constant.

In order to determine $\Omega\left(E_{k}\right)$, let us note that
for each pair $(n,l=j+\frac{1}{2})$ there are $2j+1=2l$ degenerate
states differing with values of the angular momentum projection quantum
number $m_{j}=-j,-j+1,\ldots,+j$. For a given (positive integer,
see above) $k$, the total degree of degeneracy is given by 
\begin{equation}
\sum_{l=1}^{k}\left(2l\right)=k\left(k+1\right).
\end{equation}

Therefore, the partition function $Z$ of the Dirac oscillator can
be rewritten as
\begin{equation}
Z=\sum_{k=1}^{\infty}k\left(k+1\right)e^{-\bar{\beta}\sqrt{ak+b}},\label{eq:Partition1}
\end{equation}
where $\bar{\beta}=mc^{2}\beta$, $a=4\xi$ and $b=1+2\xi$. For simplicity,
over the remainder of the work, we adopted the Natural unit system
$(\hslash=c=\kappa_{B}=1$), such that all parameters are considered dimensionless.

According to the above considerations, we can define the thermodynamic
functions of interest as follows:
\begin{align}
\bar{F} & =-\frac{1}{\bar{\beta}}\log Z,\label{freeEnergy}\\
\bar{U} & =-\frac{\partial}{\partial\bar{\beta}}\log Z,\\
\bar{S} & =\bar{\beta}^{2}\frac{\partial\bar{F}}{\partial\bar{\beta}},\\
\bar{C} & =-\bar{\beta}^{2}\frac{\partial\bar{U}}{\partial\bar{\beta}}\label{heatCapacity}.
\end{align}

As our initial evaluation, we consider the convergence of the series
in Eq. \eqref{eq:Partition1}. Applying the integral test, this series
is convergent since $f(x)=x(x+1)\exp\left(-\bar{\beta}\sqrt{ax+b}\right)$
is a monotonic decreasing function and the associated integral
\begin{eqnarray}
I(\bar{\beta}) & = & \int_{1}^{\infty}x\left(x+1\right)e^{-\bar{\beta}\sqrt{ax+b}}dx\nonumber \\
 & = & \frac{4}{a^{3}\bar{\beta}^{6}}e^{-\bar{\beta}\sqrt{a+b}}\left[60+60\sqrt{a+b}\bar{\beta}+3(11a+8b)\bar{\beta}^{2}\right.\\
 &  & \left.+\sqrt{a+b}(13a+4b)\bar{\beta}^{3}+a(4a+3b)\bar{\beta}^{4}+a^{2}\sqrt{a+b}\bar{\beta}^{5}\right],\nonumber 
\end{eqnarray}
is finite.

In order to get an insight into the convergence speed of the series,
needed to find the high-temperature limit and numerically evaluate
the partition function, we employ the Euler-Mclaurin summation formula
\cite{Abramowitz}:
\begin{eqnarray}
\sum_{n=a}^{b}f(n) & = & \int_{a}^{b}dxf(x)+\frac{1}{2}\big[f(b)+f(a)\big]-\sum_{i=2}^{p}\frac{b_{i}}{i!}\big[f^{(i-1)}(a)\big]\nonumber \\
 &  & -\int_{a}^{b}dt\frac{B_{p}\big(\{1-t\}\big)}{p!}f^{(p)}(t),
\end{eqnarray}
where $B_{p}$ and $b_{i}$ are the Bernoulli polynomials and Bernoulli
numbers, respectively, $p$ is any positive integer and the symbol $\{\cdot\}$
denotes the fractional part. After fixing a number $p=4$ of terms,
the partition function will become

\begin{eqnarray}
Z & = & \frac{240}{a^{3}\bar{\beta}^{6}}+\frac{12(a-2b)}{a^{3}\bar{\beta}^{4}}+\frac{2b(b-a)}{a^{3}\bar{\beta}^{2}}-\frac{a^{3}-6ab^{2}+4b^{3}}{12a^{3}}\nonumber \\
 & + & \frac{\bar{\beta}\left(114a^{6}+1069a^{5}b+772a^{4}b^{2}-2480a^{3}b^{3}-2560a^{2}b^{4}+768ab^{5}+1024b^{6}\right)}{6720a^{3}(a+b)^{5/2}}\nonumber \\
 & + & \frac{\bar{\beta}^{2}\left(a^{4}-10a^{3}b+20ab^{3}-10b^{4}\right)}{240a^{3}}+\mathcal{O}\left(\bar{\beta}^{3}\right).\label{eq:approx_ParticionFunction}
\end{eqnarray}

In the high-temperature regime, where $\bar{\beta}\ll1$, only the
first term in \eqref{eq:approx_ParticionFunction} have significant
contribution for the partition function. Hence, in this limit we have
\begin{equation}
Z\sim\frac{240}{a^{3}\bar{\beta}^{6}},\label{eq:partition_function_asymp}
\end{equation}
which gives the asymptotic behavior for the average energy and specific
heat: 
\begin{align}
\bar{U} & \sim 6 \tau,\label{eq:mean_energy_asymp}\\
\bar{C} & \sim 6.\label{eq:heat_capacity_asymp}
\end{align} Let us note that all functions are written in terms of the dimensionless variable 
\begin{equation}
\tau=\frac{\kappa_{B}T}{m c^{2}}\equiv\frac{T}{T_{0}},
\end{equation}where $T_{0}=\frac{m c^{2}}{\kappa_{B}}\approx5.93\times10^{9}\mbox{K}$ stands for the characteristic temperature of the system. This quantity is analogous to the so-called Debye temperature defined in solid state physics.

As in the one-dimensional case, the equations \eqref{eq:mean_energy_asymp} and \eqref{eq:heat_capacity_asymp} show that, at high temperatures, the mean energy and the specific heat  are  twice greater than the non-relativistic  three-dimensional harmonic oscillator. This result is consistent with the equipartition theory applied for an extreme relativistic ideal gas, and despite the presence of a stronger spin-orbit coupling in the Hamiltonian of the Dirac oscillator, the asymptotic behaviour of the functions described above is unaffected. Thus, at high temperatures the relativistic effects become dominant \cite{Pacheco2003}.  It is worth mentioning that this same relation holds for the one-dimensional Kemmer oscillator \cite{Boumali2007} and  the  two-dimensional noncommutative Dirac oscillator \cite{Boumali2013}.

\section{Results and discussions}

In the sequel, we briefly depict our numerical results on the evaluation  of the thermodynamic functions \eqref{freeEnergy}-\eqref{heatCapacity} and compare them with the one-dimensional case, commenting on the main features of the new profiles. Here, we plot all profiles of the thermal quantities as a function of the dimensionless temperature $\tau$ for different values of the parameter $\xi$.  Thus, we have chosen $\xi=0.10$, $0.50$, $0.75$ and $1.00$, which implies an oscillator frequency within the range $10^{19}<\omega<10^{20}$ $\mbox{Hz}$.

\begin{figure}[!htb] 
       \begin{minipage}[b]{0.48 \linewidth}
           \includegraphics[width=\linewidth]{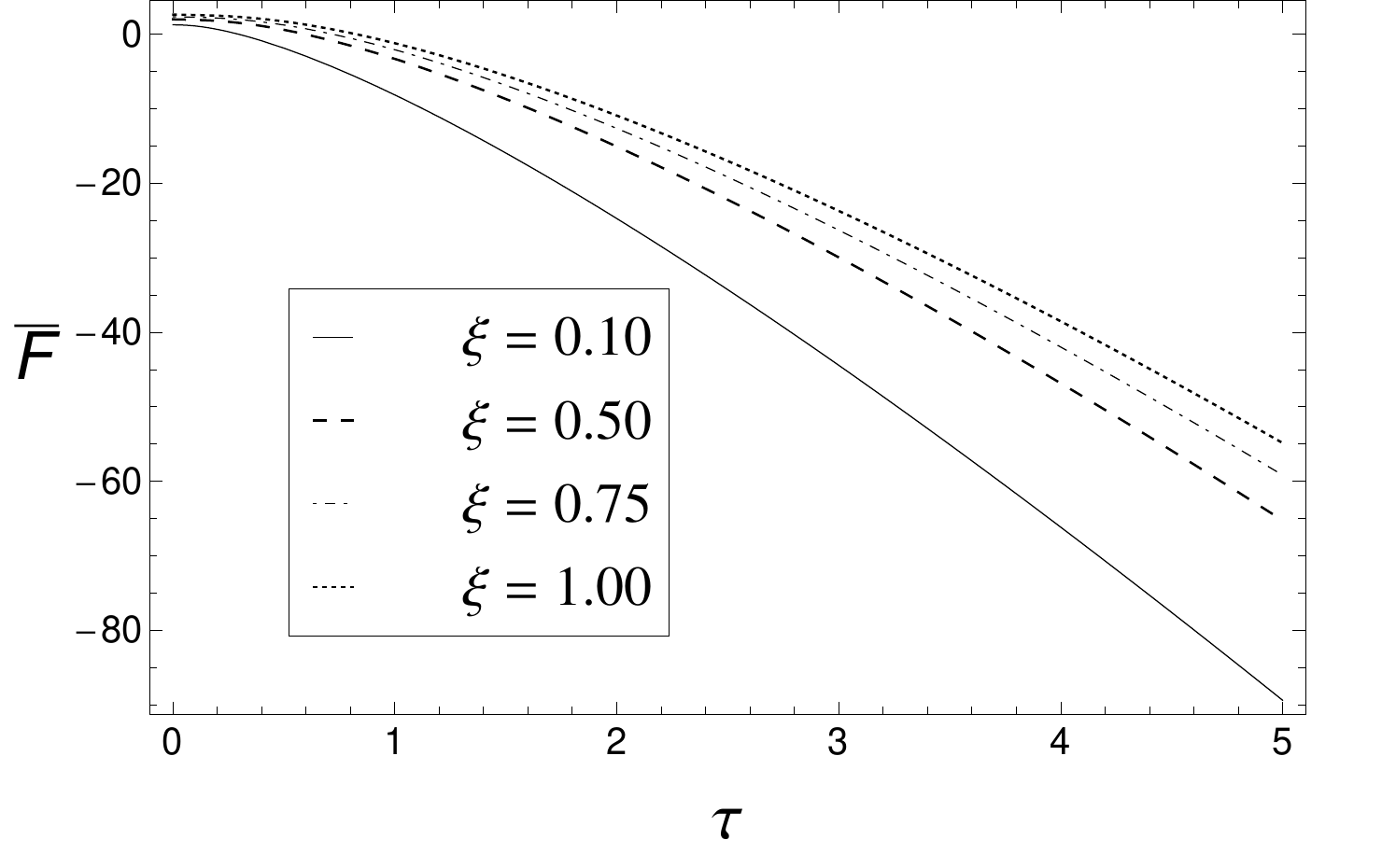}\\
           \caption{The free energy $\bar{F}$ for the 3D Dirac oscillator as a function of the dimensionless temperature $\tau=\kappa_{B}T/mc^{2}$ for different values of the parameter $\xi=\hslash\omega/mc^{2}$.}
          \label{fig:The-free-energy}
       \end{minipage}\hfill
       \begin{minipage}[b]{0.48 \linewidth}
           \includegraphics[width=\linewidth]{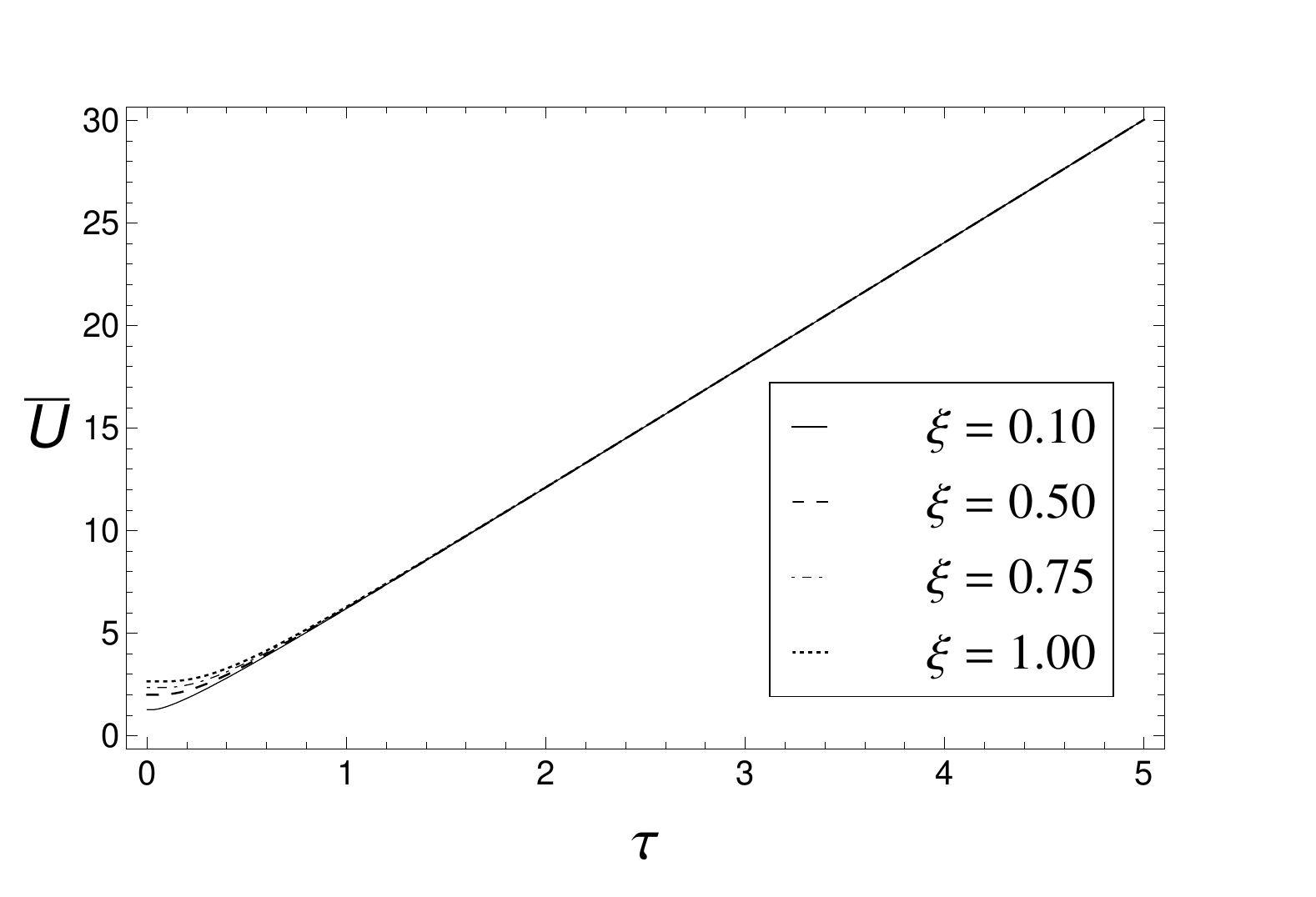}\\
           \caption{The mean energy $\bar{U}$ for the 3D Dirac oscillator as a function of the dimensionless temperature $\tau=\kappa_{B}T/mc^{2}$ for different values of the parameter $\xi=\hslash\omega/mc^{2}$.}
           \label{fig:The-mean-energy}
       \end{minipage}
   \end{figure}

\begin{figure}[!htb] 
       \begin{minipage}[b]{0.48 \linewidth}
           \includegraphics[width=\linewidth]{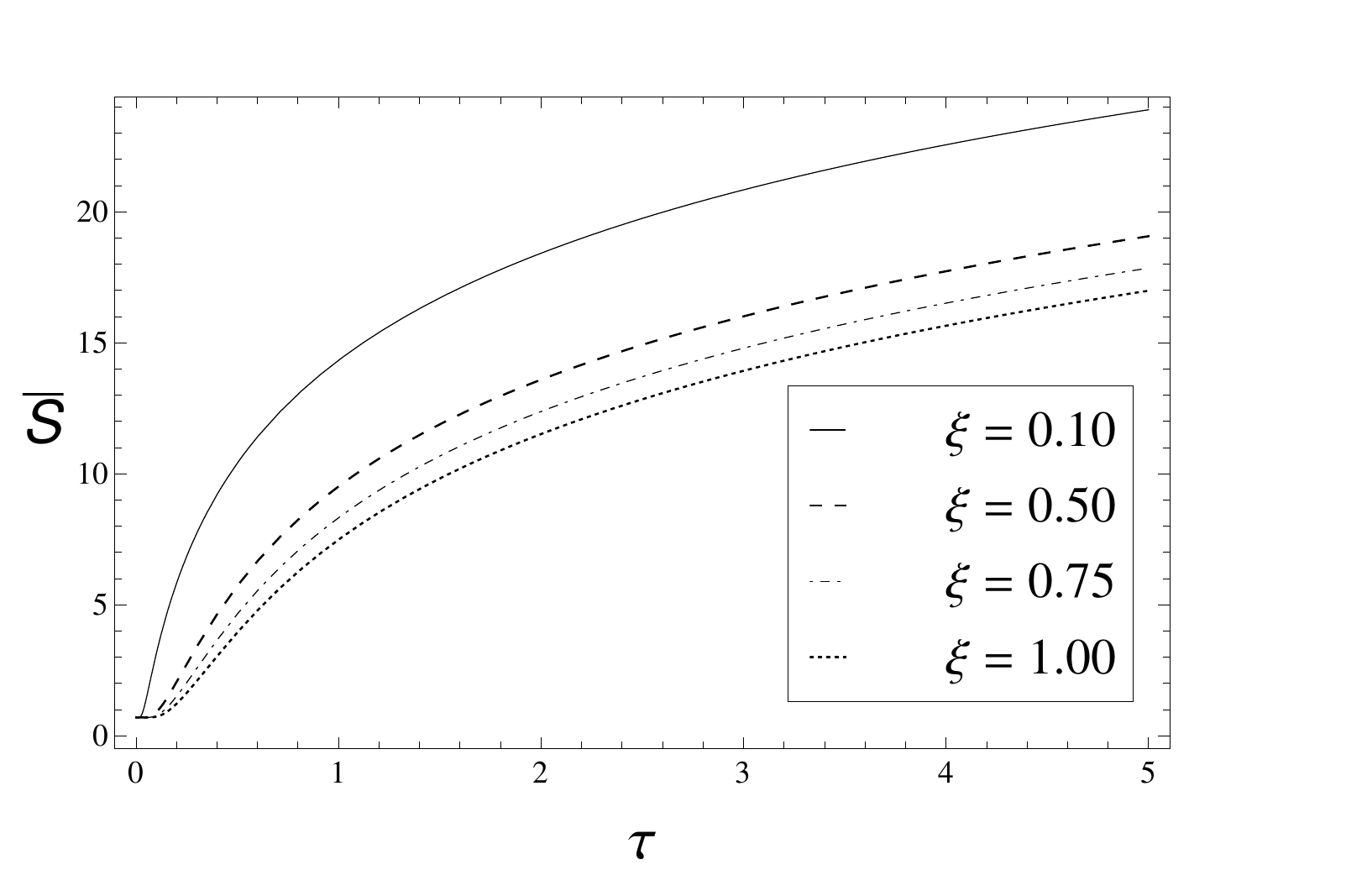}\\
           \caption{The entropy $\bar{S}$ for the 3D Dirac oscillator as a function of the dimensionless temperature $\tau=\kappa_{B}T/mc^{2}$ for different values of the parameter $\xi=\hslash\omega/mc^{2}$.}
          \label{fig:The-entropy}
       \end{minipage}\hfill
       \begin{minipage}[b]{0.48 \linewidth}
           \includegraphics[width=\linewidth]{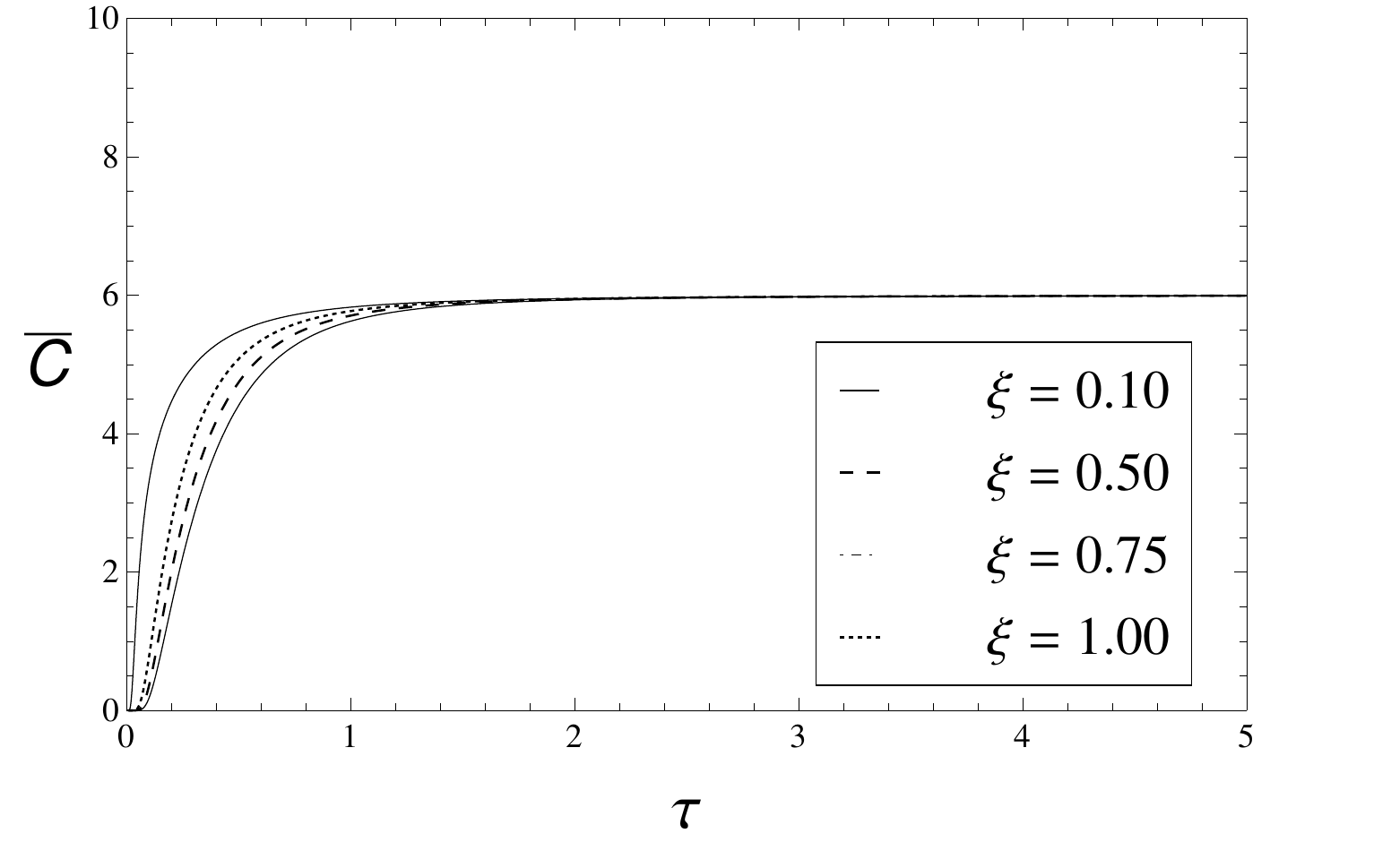}\\
           \caption{The specific heat $\bar{C}$ for the 3D Dirac oscillator as a function of the dimensionless temperature $\tau=\kappa_{B}T/mc^{2}$ for different values of the parameter $\xi=\hslash\omega/mc^{2}$.}
           \label{fig:The-specific-heat}
       \end{minipage}
   \end{figure}

The results that we obtained are shown in the figures \ref{fig:The-free-energy}, \ref{fig:The-mean-energy}, \ref{fig:The-entropy} and \ref{fig:The-specific-heat}. The overall conclusion being that the profiles of the curves behave in the same general way of their one-dimensional counterpart \cite{Pacheco2003}. However, in all cases the values ​​of the functions are raised with respect to the one-dimensional case due to the degeneracy of the energy levels. The Helmholtz free energy $\bar{F}$ is shown in Fig. \ref{fig:The-free-energy}
. Notice that for a fixed value of $\tau$, the free energy increases
when $\xi$ grows. In all cases, the profile of the curves decreases
monotonically with the temperature. In Fig. \ref{fig:The-mean-energy},
we plot the mean energy such that all curves have the same linear
behavior and exhibit very close profiles. Figure \ref{fig:The-entropy}
displays the entropy, and we identify that it increases as we reduce
the value of the parameter $\xi$. Finally, the specific heat for
the 3D Dirac oscillator is shown in Fig. \ref{fig:The-specific-heat},
which reveals that all the profiles have the same general behavior,
reaching the limit value $\bar{C}=6$ as predicted by Eq. \eqref{eq:heat_capacity_asymp},
monotonically. This result reveals that the specific heat is three times greater than that of the one-dimensional Dirac oscillator for high
temperatures \cite{Pacheco2003}. Also, we point out that this same relation holds for the one-dimensional Kemmer oscillator \cite{Boumali2007}.

\section{conclusions\label{sec:Conclusions}}

The present work was devoted to study the three-dimensional Dirac
oscillator in thermal bath. We focused on thermodynamic properties
derived from the canonical partition function at finite temperature
$T$. Indeed, the high-temperature limit has been worked out and the
asymptotic values for the mean energy and the specific heat were determined.
Also, the profiles of the main thermodynamic functions have been evaluated.
As a result, we have observed an effective increase on the values of
the functions. In particular, the asymptotic value for the specific
heat is three times bigger than the one-dimensional case. This result indicates the effect of the degeneracy on the general behaviour of the system. Recently, the first experimental realization of the Dirac oscillator was carried out by Franco-Villafa\~{n}e {\it et al}. \cite{Franco2013} and we expect that our results may be used as a good tool to study the thermodynamic properties of Dirac-like equations.

\newpage
   
\begin{acknowledgments}
The authors express their gratitude to CAPES (Coordena\c{c}\~{a}o de Aperfei\c{c}oamento de Pessoal de N\'{i}vel Superior), CNPq (Conselho Nacional de Pesquisas) and FUNCAP (Funda\c{c}\~{a}o Cearense de Apoio ao Desenvolvimento Cient\'{i}fico
e Tecnol\'{o}gico) for financial support.
\end{acknowledgments}


\end{document}